\documentclass[twocolumn,prb,aps,showpacs]{revtex4-1}

\usepackage{graphicx}
\usepackage{color}
\usepackage{dcolumn}
\usepackage{amsmath}
\usepackage{units}
\usepackage{amsmath,amssymb}
\usepackage{bibentry}

\newlength{\figwidth} 
 \newlength{\figwidthb} %
\newcommand{\CRO}{Ca$_2$RuO$_4$ } 
\newcommand{\CROns}{Ca$_2$RuO$_4$}

\begin{document}
\title{Soft spin-amplitude fluctuations in a Mott-insulating ruthenate} 

\author{A. Jain$^{1,2,\dagger}$, M. Krautloher$^{1,\dagger}$, J. Porras$^{1,\dagger}$,  G. H. Ryu$^1$, D. P. Chen$^1$, D. L. Abernathy$^3$, J. T. Park$^4$, A. Ivanov$^5$, J. Chaloupka$^6$, G. Khaliullin$^1$, B. Keimer$^{1}$, and B. J. Kim$^{1,*}$}


\address{$^1$Max Planck Institute for Solid State Research, Heisenbergstra\ss e 1, D-70569 Stuttgart, Germany} 
\address{$^2$Solid State Physics Division, Bhabha Atomic Research Centre, Mumbai 400085, India}
\address{$^3$Quantum Condensed Matter Division, Oak Ridge National Laboratory, Oak Ridge, Tennessee 37831, USA}
\address{$^4$Heinz Maier-Leibnitz Zentrum, TU M\"unchen, Lichtenbergstra\ss e 1, D-85747 Garching, Germany}
\address{$^5$Institut Laue-Langevin, 6, rue Jules Horowitz, BP 156, 38042 Grenoble Cedex 9, France}
\address{$^6$ Central European Institute of Technology,
Masaryk University, Kotl\'a\v{r}sk\'a 2, 61137 Brno, Czech Republic}
\date{\today}

\maketitle



\noindent
{\bf
Magnetism in transition-metal compounds (TMCs) has traditionally been associated with spin degrees of freedom, because the orbital magnetic moments are typically largely quenched. On the other hand, magnetic order in 4f- and 5d-electron systems arises from spin and orbital moments that are rigidly tied together by the large intra-atomic spin-orbit coupling (SOC). Using inelastic neutron scattering on the archetypal 4d-electron Mott insulator \CROns, we report a novel form of excitonic magnetism in the intermediate-strength regime of the SOC. The magnetic order is characterized by ``soft'' magnetic moments with large amplitude fluctuations manifested by an intense, low-energy excitonic mode analogous to the Higgs mode in particle physics.  This mode heralds a proximate quantum critical point separating the soft magnetic order driven by the superexchange interaction from a quantum-paramagnetic state driven by the SOC. We further show that this quantum critical point can be tuned by lattice distortions, and hence may be accessible in epitaxial thin-film structures. The unconventional spin-orbital-lattice dynamics in \CRO identifies the SOC as a novel source of quantum criticality in TMCs. 
}

\CRO is arguably best known among 4d Mott insulators\cite{Nakatsuji_1997}, particularly for its close relationship to the unconventional superconductor Sr$_2$RuO$_4$ (ref.~2)\bibentry{Maeno_1994}. Despite extensive studies over the last two decades, it remains an open question how their electronic structures evolve from one to the other\cite{Nakatsuji_2000_2,Anisimov_2002,Fang2004,Liebsch_2007,Gorelov_2010}, concomitantly with structural distortions\cite{Braden_1998,Friedt2001}. Sr$_2$RuO$_4$ adopts the tetragonal K$_2$NiF$_4$ structure with the space group $I4/mmm$. Its low-spin $d^4$ configuration, with four electrons distributed among three $t_{2g}$ levels, exhibits a superconducting ground state. It is generally believed that orbital degeneracy frustrates Mott insulation, and thus lifting the degeneracy is a vital aspect in modeling the electronic structure of \CROns. In fact, \CRO becomes a Mott insulator via a structural phase transition to the space group $Pbca$ that involves compression of the RuO$_6$ octahedra along the $c$ crystallographic axis\cite{Braden_1998,Friedt2001}.

\begin{figure*}
\centerline{\includegraphics[width=1.5\columnwidth,angle=0]{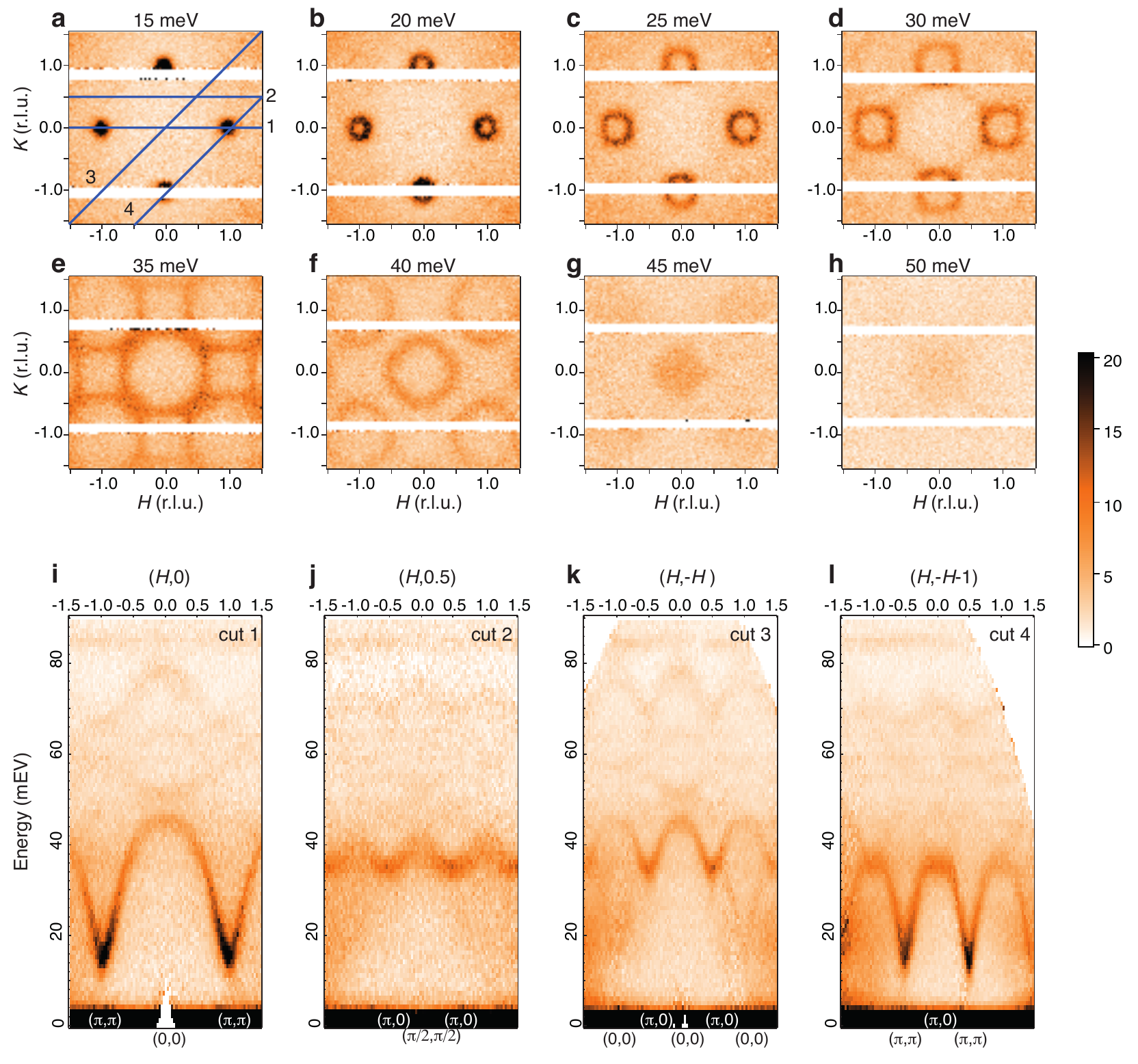}}
\caption{{\bf Time-of-flight INS spectra of \CROns.}  (a-h) Constant energy maps of INS intensity in the ($H$,$K$) plane with $L$ integrated over the range 0\,$\leq$\,$L$\,$\leq$6. The wave-vectors are expressed in reciprocal lattice units (r.l.u.) of a tetragonal cell with  $a^{*}$=\,$b^{*}$=\,1.16\,\AA$^{-1}$ and $c^{*}$=\,0.53\,\AA$^{-1}$. At $E$\,=\,15\,meV, intense magnons are observed at the reduced wave-vectors $(\pm 1,0)$ and $(0,\pm 1)$, which correspond to the antiferromagnetic ordering wave-vector $(\pi,\pi)$ for the 2D square lattice Brillouin-zone.  (i-l) Energy spectra along high-symmetry directions as shown in blue lines in panel (a). The intensity is in arbitrary units. }\label{fig:fig1}
\end{figure*}

First-principles calculations\cite{Fang2004,Gorelov_2010} estimate the tetragonal crystal-field splitting $\Delta$\,$\sim$\,0.3\,eV, which well exceeds the SOC $\xi$\,$\sim$\,0.15\,eV (ref.~10)\bibentry{Mizokawa_2001} and hence largely quenches the orbital angular momentum. With two electrons occupying the $xy$ orbital stabilized by the compressive tetragonal distortion, the remaining orbitals $yz$ and $zx$ become half-filled, and thus a Heisenberg antiferromagnet (AF) would be expected from superexchange interactions among spin-only moments.  Contrary to this widely accepted picture, our comprehensive set of inelastic neutron scattering (INS) data reveals a magnetic phase distinct from a conventional Heisenberg AF, implying a nontrivial interplay\cite{Khaliullin_2013} among SOC, exchange interactions, and orbital-lattice coupling in \CROns.

\begin{figure}
\centerline{\includegraphics[width=0.9\columnwidth,angle=0]{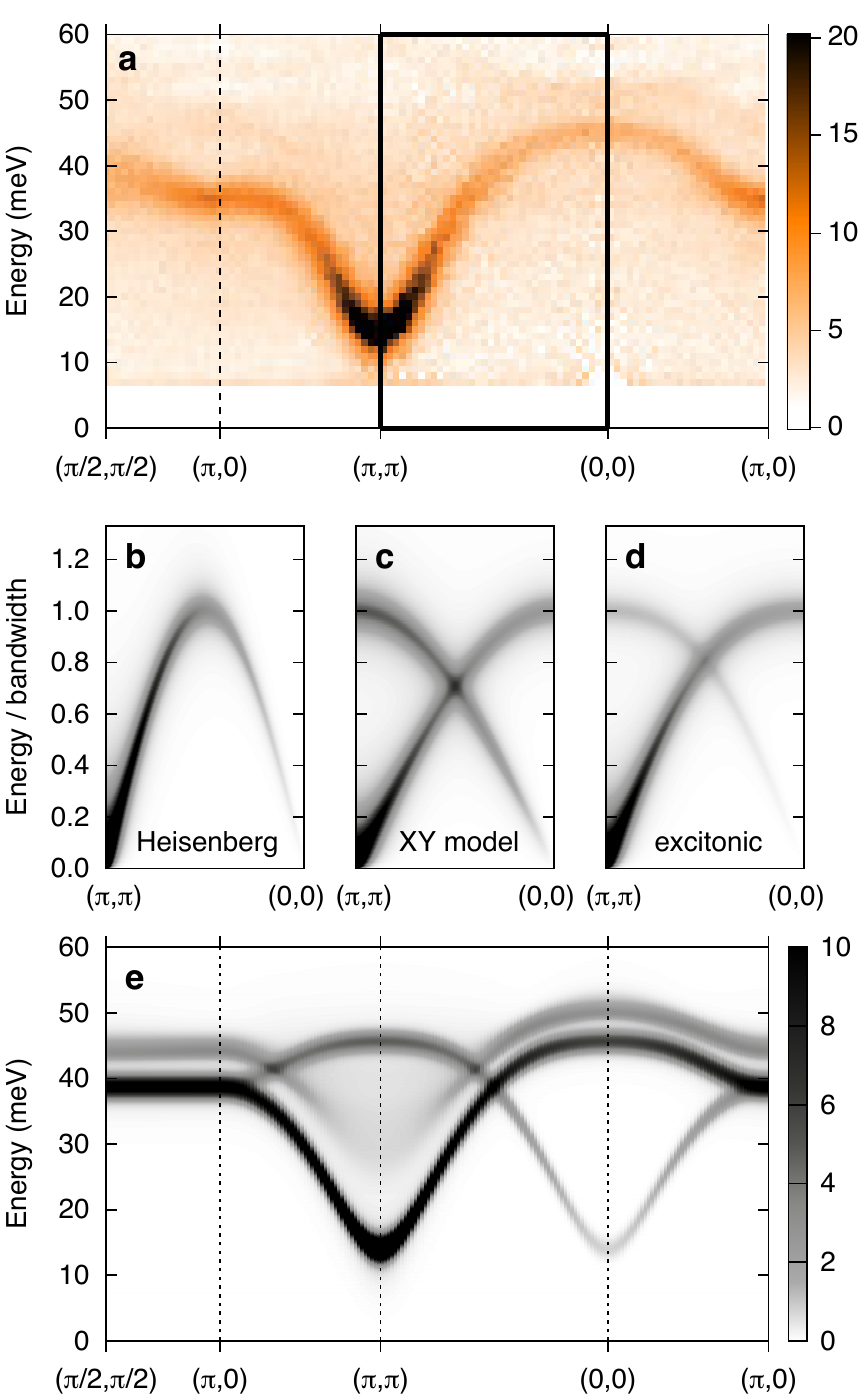}}
\caption{{\bf Spin-wave dispersion strongly deviating from the Heisenberg model.} (a) TOF INS spectra along high symmetry directions. The intensity is in arbitrary units. The thick box indicates the region to be compared with theory simulations in panels b-d. Textbook spin-wave spectra for (b) a Heisenberg and (c) an XY antiferromagnet. (d) Two transverse modes in the excitonic model. (e) The full spectra of the excitonic model including the amplitude mode along high symmetry directions. The parameters used in eq.~(1) were $E$\,$\simeq$\,24, $J$\,$\simeq$\,5.6, $\epsilon$\,$\simeq$\,4.0\,meV, and $\alpha$\,=\,0. The decay of the amplitude mode into the two-magnon continuum is modeled in the Supplementary Information. }\label{fig:fig1}
\end{figure}

\begin{figure}
\centerline{\includegraphics[width=0.9\columnwidth,angle=0]{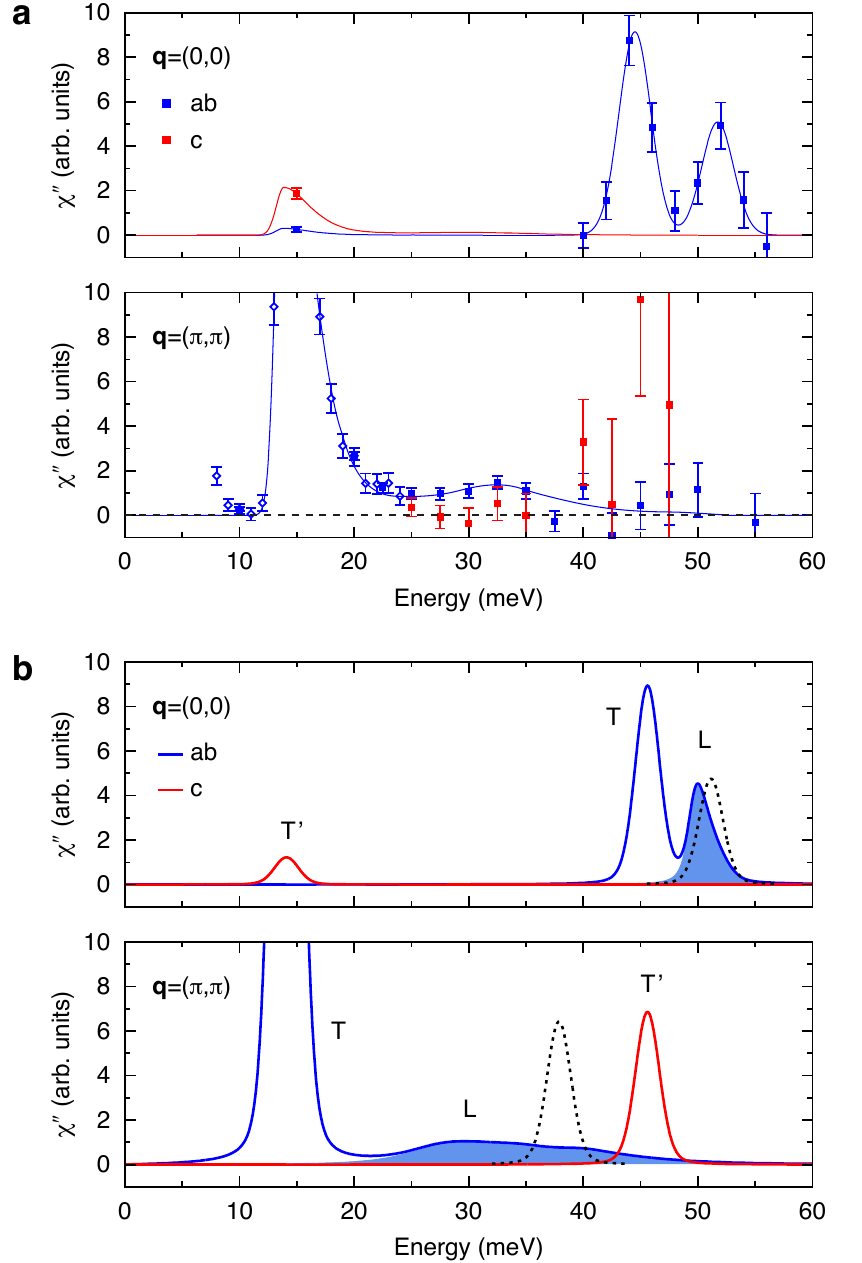}}
\caption{{\bf Identification of the magnetic modes with polarized INS and their comparison to model calculation.} (a) Imaginary part of the dynamic spin susceptibility obtained by normalising the INS spectra measured at $\mathbf{q}$\,=\,(0,0) (Supplementary Fig.~3) and $\mathbf{q}$\,=\,($\pi$,$\pi$) (Supplementary Fig.~4) with respect to the orientation factor and the isotropic form factor for Ru$^{+}$ (Supplementary Fig.~2). Because our sample mosaic is twinned with mixed $a$ and $b$ axes, we are able to distinguish only between in-plane (blue symbols and solid lines) and out-of-plane (red symbols and solid lines) polarizations of the magnetic modes. Solid symbols show data with the background removed by taking the difference between two spin-flip channels, and open symbols show data from a single spin-flip channel (see Supplementary Information). Solid lines are guides to the eye. Error bars denote one standard deviation. The lineshape and width of the folded mode at $E$\,=\,15 meV were constrained to be the same as the main transverse mode. (b) Model calculation with the parameters as given in the caption of Fig.~2. ``T'' (``L'') stands for transverse (amplitude) modes. Black dashed lines show the bare ``L" modes with the decay mechanism into two ``T'' magnons turned off. }\label{fig:fig1}
\end{figure}

Figures 1a-h and 1i-l exhibit constant-energy maps and energy spectra along high-symmetry directions measured by time-of-flight (TOF) INS, respectively, comprising spin-wave dispersions in the energy range 14\,$\lesssim$\,$\hbar\omega$\,$\lesssim$\,45 meV and phonon modes above $\sim$50\,meV. The magnetic nature of the former is explicitly confirmed by using spin-polarized neutrons (to be presented later in Fig.~3), and the non-magnetic nature of the latter is inferred from exhaustion of all magnetic modes and also through comparison with the known phonon modes\cite{Rho2003}.

The spin-wave dispersions along all high-symmetry directions are summarized in Fig.~2a.
The spin-waves emanate from the ordering vector ($\pi$,$\pi$) with a gap of $\approx$14\,meV, and reaches their maximum energy at $\mathbf{q}$\,=\,(0,0) (see also  Fig.~1g), qualitatively deviating from the well-known Heisenberg AF magnon dispersion (Fig.~2b). This is a clear manifestation of unquenched orbital magnetism, 
which, on a phenomenological level, can be accounted for by 
the effective $S$\,=\,1 Hamiltonian:
\begin{equation}
H = J\sum_{\langle ij\rangle} (\mathbf{S_{i}\cdot S_{j}}- \alpha S_{zi}S_{zj})
+ E \sum_i S^2_{zi} +\epsilon \sum_i S^2_{xi}.
\end{equation}
This minimal model 
includes single-ion 
anisotropy ($E$ and $\epsilon$) terms induced by tetragonal ($z$\,$\parallel$\,$c$) and 
orthorhombic ($x$\,$\parallel$\,$a$) distortions, correspondingly, as well 
as the XY-type exchange anisotropy ($\alpha>0$). Experimentally, the magnetic moment is known to lie along the $b$-axis\cite{Braden_1998}.

Formally, one can account for the monotonic upward dispersion from 
($\pi$,$\pi$) to (0,0) by tuning $\alpha$ from the Heisenberg limit 
($\alpha$\,=\,0) to the XY-limit ($\alpha$\,=\,1) 
without invoking the single-ion anisotropy.  However, also expected in the XY model is an intense folded mode with an equal intensity at the crossing point ($\frac{\pi}{2}$,$\frac{\pi}{2}$) (Fig.~2c), which is absent in the experimental data. The failure of the XY model is actually expected since the magnetic anisotropy in $S$\,$\geq$\,1 systems is generally dominated by the single-ion terms. Among them, we expect $E$ to be much
larger than $\epsilon$ based on the crystal structure. Indeed, tuning the parameters to large $E$, now with $\alpha$\,$\approx$\,0, reproduces the XY-like dispersion with the intensity of the folded mode ($I'$) much weaker than that of the main branch ($I$) (Fig.~2d); their ratio is given by 
$I'/I$\,$\simeq$\,$(\tau-1)/(\tau+1)$ where 
$\tau$\,=\,$J/J_{cr}$\,$\approx$\,$8J/E$ 
(see Supplementary Information). We shall see later that $E$ is directly given by the energy scale of the SOC, which in a solid is effectively reduced by the tetragonal distortion. 

\begin{figure}
\centerline{\includegraphics[width=0.8\columnwidth,angle=0]{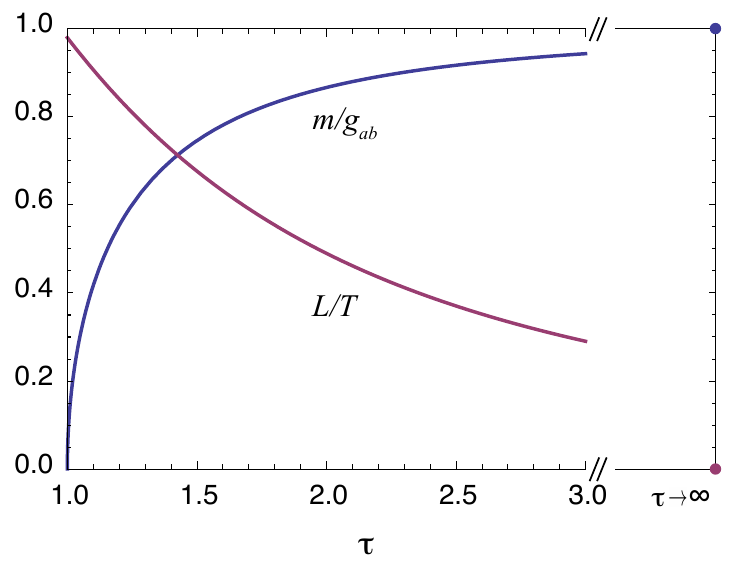}}
\caption{{\bf Quantification of the proximity to the QCP.} Evolution of the intensity ratio between the amplitude (L) and the transverse (T) modes at $\mathbf{q}$\,=\,(0,0), and the static magnetic moment (m) (normalized by the $ab$ -plane $g$-factor, see Fig.~S1) as a function of $\tau$\,=\,$J/J_\mathrm{cr}$ (see text). $\tau$\,=\,1 at the QCP.}\label{fig:fig1}
\end{figure}

In the limit of very large $E$, the ground state becomes a non-magnetic $S_z$\,=\,0 singlet (i.e. no Kramers degeneracy) with spins disordered in the $xy$ plane---known as `spin nematic' phase\cite{Podolsky2005}. As $E$ is reduced, the $J$-interaction 
induces magnetic order when $\tau$\,=\,1, with magnon dispersions mimicking that of the XY model. However, the static moment can be small compared to its saturation value (2$\mu_B$), vanishing continuously at the quantum critical point (QCP)---a key distinction from the XY model with rigid spins of fixed length. This may account for the large reduction (35\%) of the observed moment in \CRO (ref.~8). A hallmark of this soft magnetism is the existence of a well-defined, low-energy amplitude mode---a condensed matter analog of the Higgs particle\cite{Ruegg_2008,Endres_2012,Sachdev_2011}. 

For quantitative verification of this scenario, we calculate the spin-wave spectra using eq.~(1) (see Supplementary Information). 
Although the full model may include more terms, such as the Dzyaloshinskii-Moriya
interaction due to rotation and tilting of the RuO$_6$ octahedra, pseudodipolar interactions present whenever SOC is nonzero, and biquadratic exchange, we find that our minimal model captures all salient features of the data.

Near the QCP ($\tau$\,$\sim$\,1), the textbook spin-wave theory with rigid spin-length has to be modified (see, e.g., refs.~17,18)\bibentry{Matsumoto2004,Sommer2001}. A modified spin-wave theory gives the spectra shown in Fig.~2e. The parameters in eq.~(1) were determined by fitting the dispersion of the main transverse mode to the observed one. These parameters are also well justified from a microscopic theory (Supplementary Fig.~1). In addition to the main and folded transverse modes, our model calculation predicts an amplitude mode of sizable intensity, well defined in most of the Brillouin zone away from ($\pi$,$\pi$). In other words, the form  of the dispersion of the transverse modes already indicates soft magnetism with an unsaturated moment, which in turn gives rise to an intense amplitude mode. Indeed, a vague hint of the amplitude mode is already visible in the TOF INS spectra (Fig.~2a). 

For an unambiguous identification of all three magnetic modes, we performed polarized INS using the standard XYZ-difference method. Fig.~3a plots the imaginary part of the dynamic spin susceptibility extracted from the INS data, to be compared with the calculated one plotted in Fig.~3b. At $\mathbf{q}$\,=\,(0,0), we observe the folded mode at $\approx$\,14 meV, which is not clearly seen in the TOF INS spectra because of its $c$-axis polarized nature; its intensity is largely suppressed by the INS orientation factor for the scattering geometry used with the TOF spectrometer. At $\mathbf{q}$\,=\,($\pi$,$\pi$), a quantitative comparison of the folded mode, expected at $\approx$\,45\,meV, is difficult because of insufficient statistics, but the intensity deviating from zero is statistically significant.

Next, we turn to the in-plane response. At $\mathbf{q}$\,=\,(0,0), in addition to the large peak of the main transverse mode at $\approx$\,45\,meV, we observe a clear peak at $\approx$\,52\,meV, where the amplitude mode is expected (Fig.~2e). Because the other two transverse modes have been unambiguously identified, this peak must be the amplitude mode. Its magnetic and in-plane polarized nature, together with the excellent match of its energy position and spectral weight with the model calculation, fully confirms this assignment.

Near $\mathbf{q}$\,=\,($\pi$,$\pi$), the amplitude mode is unstable against decaying into two Goldstone magnons, especially in a two-dimensional lattice in which the longitudinal susceptibility has an infrared singularity\cite{Podolsky_2011}. In our case, the singularity is cut off by the nonzero gap ($\approx$\,14\,meV) of the transverse mode, and thus the amplitude mode is expected to be visible, albeit with very broad lineshape. In the experimental spectrum (Fig.~3a), we observe intensities well outside of the error bar from zero, distributed in the energy range 20\,--\,45\,meV, in excellent accord with the calculated spectrum (Fig.~3b). 

\begin{figure}
\centerline{\includegraphics[width=1\columnwidth,angle=0]{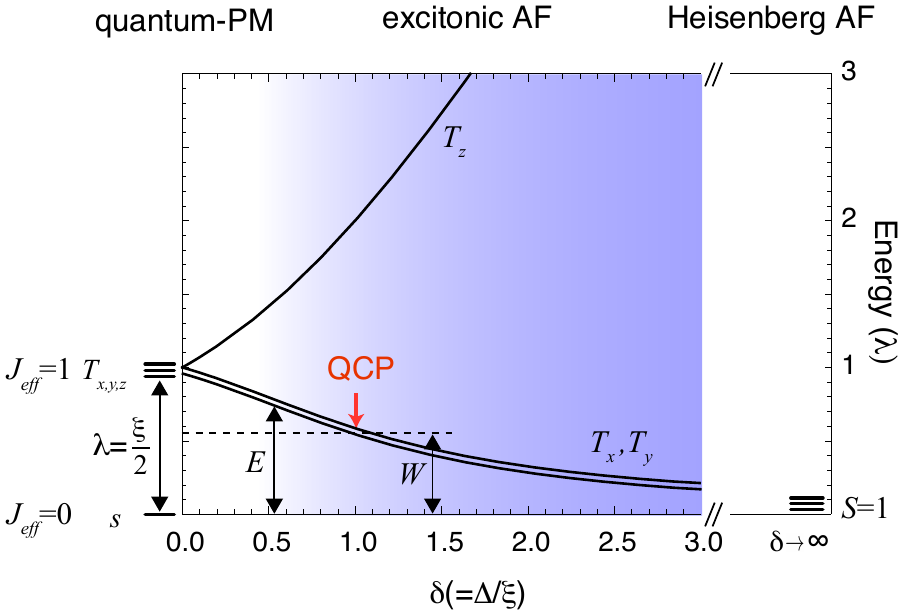}}
\caption{{\bf Microscopic mechanism of the QCP driven by tetragonal lattice distortion.} Crystal-field splitting of the $J_{\mathrm {eff}}$\,=\,1 triplet effectively lowers the energy scale of SOC from $\lambda$ to $E$. The QCP occurs when $E$ becomes equal to the strength of the exchange field $W$\,$\approx$\,2$zJ$. From the measured value of intensity ratio $L/T$ (Fig.~4), we estimate the lower bound for $\delta$ to be about 1.0 for \CRO (see Supplementary Information). The blue shading indicates the region where the effective $S$\,=\,1 model is valid.}\label{fig:fig1}
\end{figure}

In contrast, the amplitude mode at $\mathbf{q}$\,=\,(0,0) maintains its integrity, because kinematic restrictions suppress its decay into the two-magnon continuum. This allows us to quantify the proximity of the system to the QCP by measuring the intensity of the amplitude mode relative to that of the transverse mode. At the QCP ($\tau$\,=\,1), where the distinction between transverse and amplitude modes vanishes, their ratio at $\mathbf{q}$\,=\,(0,0) is $\simeq$\,1 (equality holds when the gap is zero), and approaches zero as the moment saturates (Fig.~4). The measured intensity ratio of 0.55\,$\pm$\,0.11 translates to $\tau$\,$\approx$\,1.8. In principle, the size of the static moment contains the same information, but only after corrections due to $g$-factors (Supplementary Fig.~1), covalency, and quantum fluctuations, have been properly taken into account, which are model-dependent and fraught with systematic uncertainties. 

Having established the existence of the amplitude mode characterizing the excitonic magnetism in \CROns, we now turn to the microscopic origin of the phenomenological model. Obviously, the qualitative departure from the conventional Heisenberg magnetism is a manifestation of strong SOC, which was, in fact, indicated in earlier experiments\cite{Mizokawa_2001,Haverkort_2008,Fatuzzo_2015}, although neglected in most theories\cite{Anisimov_2002,Hotta_2001,Liebsch_2007,Gorelov_2010}. For the low-spin Ru$^{4+}$ ion, the SOC ($\xi$) in cubic symmetry leads to a $J_{\mathrm {eff}}$\,=\,0 singlet and a $J_{\mathrm {eff}}$\,=\,1 triplet, the former being the ground state and the latter having an excitation energy of $\lambda$\,=\,$\xi/2$. While the ground state is thus expected to be nonmagnetic in the strong SOC limit, a quantum phase transition to a magnetic phase can occur with condensation of one of the $J_{\mathrm{eff}}$\,=\,1 triplet when the exchange coupling is strong enough to overcome the SOC\cite{Khaliullin_2013}. 


Considering the free-ion value of $\lambda$\,$\simeq$\,75\,meV, a quick inspection shows that $J$(\,$\sim$\,5.6 meV) is well below $J_{cr}$\,($\sim$\,$\lambda/2z$\,$\sim$\,9.4 meV, where $z$\,=\,4 for a square lattice), and thus \CRO would be a quantum paramagnet. We note that neither $J$ nor $\lambda$ is easily tunable. However, the unique energy hierarchy in 4$d$ TMCs with $\Delta$\,$\sim$\,$\xi$ allows tuning them across the QCP. Figure 5 illustrates this point; $\Delta$ splits the $J_{\mathrm {eff}}$\,=\,1 triplet into a doublet ($T_x$,$T_y$) and a singlet ($T_z$), and thereby lowers the energy cost to excite a triplon. Thus, for fixed $J$ and $\lambda$, the magnetic QCP is driven by $\Delta$, which can be tuned through epitaxial strain. In the limit $\delta$\,$\rightarrow$\,$\infty$, the doublet merges with the $J_{\mathrm {eff}}$\,=\,0 singlet, and the resulting three levels constitute an isotropic $S$\,=\,1 Heisenberg AF (Fig.~5). For most of the phase diagram between the cubic and Heisenberg limits, the $T_z$ level is very high in energy and thus can be neglected (shaded region in Fig.~5), so that the remaining three levels $\{s,T_x,T_y\}$ form the basis for our effective $S$\,=\,1 Hamiltonian of Eq.~(1).  From this energy diagram, it is clear that $E$ is nothing but the energy scale of
$\lambda$, effectively lowered by $\Delta$. For $S$\,=\,$\frac{1}{2}$ systems (e.g. iridates),  this single-ion anisotropy term is forbidden even with large SOC, and the new physics reported herein is special to the $S$\,=\,1 systems.

Conceptually, one may find a useful analogy to the magnetic field driven QCPs in dimer systems, such as TlCuCl$_3$ (refs.~14,23,24).\bibentry{Ruegg_2003,Ruegg_2008,Ruegg_2008_2} In these systems, a similar singlet-triplet structure is realized in pairs of spins on each dimer.  In the ruthenates, however, the energy scale characterizing the magnetic dynamics is much larger than in the dimer systems. Indeed, the characteristic energy is comparable to room temperature, which implies robust quantum effects that may be exploited for practical applications. Although such a large energy scale QCP cannot be tuned by laboratory-scale magnetic fields, ruthenates offer an alternative means of tuning via crystalline electric fields in epitaxial thin-film structures\cite{Wu2013}. Moreover, the singlet-triplet structure realized within a single ion removes the stringent requirement of a dimer structure in the crystal lattice and thus has a wider applicability for transition-metal compounds containing 4d and 5d ions\cite{Khaliullin_2013,Cao2014,Meetei_2015}.
  
We remark that the distinct SOC physics demonstrated herein is not limited to \CROns, but should be widely visible in other Ru$^{4+}$ compounds and other 4d-electron systems. Given that interpretations of 4d TMCs have largely been based on models neglecting the vital role of the SOC, our work calls for reinvestigation of the distinct correlated electron phenomena in 4d TMCs, such as the purported orbital ordering transition in La$_4$Ru$_2$O$_{10}$ (ref.~28)\bibentry{Khalifah_2002}, dimer lattice formation in Li$_2$RuO$_3$ (ref.~29)\bibentry{Miura2007}, and the spin-one Haldane gap in Tl$_2$Ru$_2$O$_7$ (ref.~30)\bibentry{Lee_2006}.

\vspace {20 pt}
\noindent
{\bf Acknowledgements} We acknowledge financial support from the German Science Foundation (DFG) via the coordinated research program SFB-TRR80, and from the European Research Council via Advanced Grant 669550 (Com4Com). The experiments at Oak Ridge National Laboratory's Spallation Neutron Source were sponsored by the Division of Scientific User Facilities, US DOE Office of Basic Energy Sciences. J.C. was supported by GACR (project no. 15-14523Y) and by ERDF under project CEITEC (CZ.1.05/1.1.00/02.0068).

\vspace{20 pt}
\noindent
{\bf Methods}

\vspace{10 pt}
\noindent
{\bf Sample synthesis \& characterization}
Single crystals of \CRO were grown by the floating zone method with RuO$_{2}$ self-flux\cite{Nakatsuji2001}. The lattice parameters $a$\,=\,5.415\,\AA, $b$\,=\,5.5512\,\AA, and $c$\,=\,11.944\,\AA\,  were determined by x-ray powder diffraction, in good agreement with the parameters reported in the literature\cite{Braden_1998} for the ``S'' phase with short $c$-axis lattice parameter. The magnetic ordering temperature $T_\mathrm{N}$\,=\,110\,K was determined using magnetization measurements in a Quantum Design SQUID-VSM device.

\vspace{10 pt}
\noindent
{\bf Time-of-flight inelastic neutron scattering}
For the TOF measurements, we co-aligned about 100 single crystals with a total mass of $\sim$\,1.5\,g into a mosaic on Al plates. Approximately half of the crystals were rotated 90$^\circ$ about the $c$-axis from the other half (Supplementary Fig.~2). The in-plane and $c$-axis mosaicity of the aligned crystal assembly were  $\lesssim$\,3.2$^\circ$ and  $\lesssim$\,2.7$^\circ$, respectively. The measurements were performed on the ARCS time-of-flight chopper spectrometer at the Spallation Neutron Source, Oak Ridge National Laboratory, Tennessee, USA. The incident neutron energy was 100~meV. The Fermi chopper and  $T_{0}$ chopper frequencies were set to 600 and 90 Hz, respectively, to optimize the neutron flux and energy-resolution. The measurements were carried out at {\it T}\,=\,5\,K. The sample was mounted with ($H$,0,$L$) plane horizontal. The sample was rotated over 90$^\circ$  about the vertical $c$-axis with a step size of 1$^\circ$. At each step data were recorded over a deposited proton charge of 3 Coulombs ($\sim$\,45 minutes) and then converted into 4D $S(\mathbf {Q},\omega)$ using the HORACE software package\cite{Horace} and normalized using a vanadium calibration.

\vspace{10 pt}
\noindent
{\bf Polarized inelastic neutron scattering}
Preliminary triple-axis measurements, in order to reproduce the TOF results and determine the feasibility of the polarized experiment, were done in the thermal three axes spectrometer PUMA at the FRM-II, Garching, Germany. The measurements were done on the same sample used for the TOF experiment. To optimize the flux and energy resolution, double-focused PG (002) and Cu (220) monochromators, for measurements below and above 30 meV respectively, and a double-focused PG (002) analyzer were used, keeping $k_{\mathrm f}$\,=\,2.662\,\AA$^{-1}$ constant.
For the polarized triple axis measurement we remounted the crystals from the TOF experiment on Si plates and increased the number of crystals to obtain a total sample mass of $\sim$\,3\,g. The mosaicity of this sample was  $\lesssim$\,3.2$^\circ$ and  $\lesssim$\,2.6$^\circ$ for in-plane and $c$-axis, respectively. The experiment was performed on the IN20 three-axis-spectrometer at the Institute Laue-Langevin, Grenoble, France. For the XYZ polarization analysis, we used a Heusler (111) monochromator and analyzer in combination with Helmholtz coils at the sample position. Throughout the experiment we used a fixed $k_{\mathrm f}$\,=\,2.662\,\AA$^{-1}$ and performed polarization analysis along energy and $H$ scans at ($\pi$,$\pi$) and (0,0), keeping $L$ as small as possible constrained by kinematic restrictions.
 
\bibliography{biblio.bib}


\newpage
\noindent
{\bf Supplementary Information}

\newcommand{\vc}[1]{\boldsymbol{#1}}

\vspace{20 pt}
\noindent
{\bf A. Microscopic model}

\vspace{10 pt}
\noindent
Using standard second order perturbation theory, we calculated the magnetic exchange coupling fully incorporating the tetragonal crystal field splitting $\Delta$, atomic SOC constant $\xi$, and Hund's coupling $\eta$ measured in units of the Coulomb interaction $U$. In terms of the original spin and orbital 
moments $\lvert S_M, L_M\rangle$, the wave functions for the basis states $\{s,T_x,T_y\}$ with 
$T_x=\frac1{i\sqrt2}(T_1-T_{-1})$, $T_y=\frac1{\sqrt2}(T_1+T_{-1})$ are given by 
\begin{align}
&|s\rangle = \sin\theta_0 \,\tfrac1{\sqrt2}(|1,-1\rangle+|-1,1\rangle)
           -\cos\theta_0 \,|0,0\rangle \;, \\
&|T_{+1}\rangle = \cos\theta_1 |1,0\rangle - \sin\theta_1 |0,1\rangle \;, \\
&|T_{-1}\rangle = \sin\theta_1 |0,-1\rangle - \cos\theta_1 |-1,0\rangle.
\end{align}
Here the angles $\theta_0$, $\theta_1$ are defined through
\begin{equation}
\tan\theta_1 = \frac1{\delta+\sqrt{1+\delta^2}} \;, \quad
\tan\theta_0 = \sqrt{1+\beta^2}-\beta, 
\end{equation}
where $\delta$\,=\,$\Delta/\xi$ and $\beta=\frac1{\sqrt2}(\delta-\frac12)$. The above wave functions form a basis for the effective spin-one model used in the main text.   

The energy $E$ in eq.~(1), which gives a singlet-doublet splitting in Fig.~5, is
\begin{equation}
E=\frac{\xi}{2} \left(\frac{\sqrt{2}}{\beta+\sqrt{1+\beta^2}}-\frac{1}{\delta+\sqrt{1+\delta^2}}\right).
\end{equation}




Figure S1a shows the exchange constant $J$ and the anisotropy $\alpha$ 
for $\eta$\,=\,0 and $\eta$\,=\,0.2; the latter would be more realistic. We note that the effective 
$S$\,=\,1 model is valid only 
for $\delta$\,$\gtrsim$\,0.5 when $T_z$ can be neglected; closer to the cubic limit ($\delta$\,$\sim$\,0), one has to use 
the full singlet-triplet model Hamiltonian of ref.~11. The calculation shows
that $\alpha$ is always small in comparison to $J$, insensitive to the value
of $\eta$, in the entire range of $\delta$ where the model is valid,
confirming that the anisotropy due to two-ion exchange is small. 

The value of $J$\,=\,5.6\,meV used to fit the spin-wave dispersion returns a
reasonable lower bound for $4t^2/U$\,$\sim$\,30\,$\mathrm{meV}$, comparable to
that found in 
$t_{2g}$ orbital vanadates (ref.~${\mathrm{S1}}$). For this estimate, we used $\eta$\,=\,0.2, which reduces $J$ from its nominal value to about $J$\,$\simeq$\,0.8\,$t^2/U$. The actual value of $4t^2/U$ may be higher than the above estimate because the magnon bandwidth (and hence $J$) can be reduced by magnon softening effects in orbitally degenerate systems$^{\mathrm{S2}}$; in the present case, coupling to the higher-lying $J_{\mathrm{eff}}$\,=\,2 manifold may result in a sizable magnon renormalization. 

Next, we provide a realistic range of $\delta$ for \CROns. The measured intensity ratio of $L/T$ (Fig.~4) translates to $\tau$\,=\,1.5\,--\,2.2, which in turn translates to $E$\,=\,$21$\,--\,$29$ meV. Taking $\lambda$\,=\,75\,meV at its ionic value, $\delta$ is estimated from Fig.~5 to be in the range $1.6$\,--\,$2.3$. However, the covalency effect may substantially reduce the SOC in a solid$^{\mathrm{S3}}$; taking, e.g., the covalency factor $\kappa$\,=\,2/3, we would find $\lambda$\,=\,50\,meV and $\delta$ in the range of 1.0-1.6. These conservative estimates 
show that \CRO is well inside the region where the model in eq.~(1) is valid. We further provide the $g$-factors in Fig.~S1b calculated from the matrix elements of the magnetic moment  
$\vc m$\,=\,$2$\,$\vc S$\,+\,$\kappa$\,$\vc L$, including the covalency factor ($\kappa$).

\begin{figure}
\includegraphics[width=0.85\columnwidth]{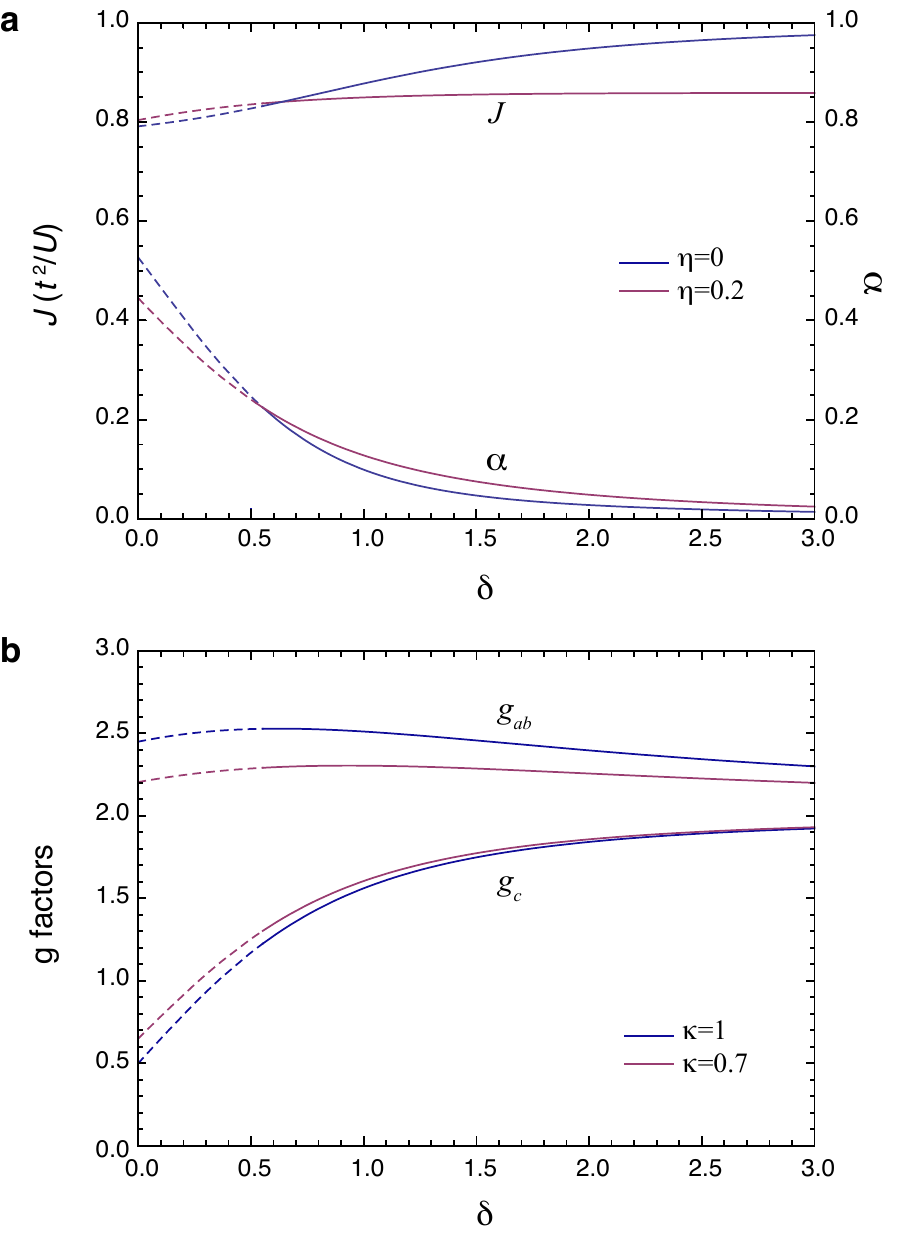}
\setcounter{figure}{0}
\renewcommand{\thefigure}{S\arabic{figure}}
\caption{{\bf Coupling constants and $\mathbf {g}$-factors from a microscopic model.} (a) $J$ and $\alpha$ (see eq.~1) as a function of $\delta$. (b) $g$-factors reduced by covalency ($\kappa$). The effective 
$S$\,=\,1 model is only valid for $\delta$\,$\gtrsim$\,0.5 (solid lines).}\label{fig:fig1}
\end{figure}

\vspace{10 pt}
\noindent
{\bf B. Mode dispersions and intensities}

\vspace{10 pt}
\noindent
The excitation spectra for the model in eq.~(1) formulated in the basis
$\{s,T_x,T_y\}$ were calculated using the modified spin-wave theory for
the models where the QCP is associated with triplet 
condensation (see refs.~11,17,18).
The energy and magnetic intensity of the longitudinal (amplitude) mode 
obtained within the harmonic approximation reads as
\begin{equation}
\omega_{L\vc q} = W\sqrt{1+\frac{\gamma_{\vc q}}{\tau^2}} \;,\quad
I_{L\vc q} \propto \frac{g_{ab}^2}{\tau}\frac1{\sqrt{\tau^2+\gamma_{\vc q}}} \;,
\end{equation}
where $W$\,=\,$8J$ is the energy scale. This mode is most dispersive and intense
near the QCP ($\tau$\,$\sim$\,1) while in the rigid-spin limit
($\tau$\,$\gg$\,1), it 
flattens and vanishes. To describe the main ($T$) and folded ($T'$)
transverse modes, we introduce two auxilliary quantities
\begin{align}
a_{\vc q}&=\tfrac12 W(1+\tfrac1{\tau})(1+\gamma_{\vc q})+\epsilon\;,\notag\\
b_{\vc q}&=\tfrac12 W(1+\tfrac1{\tau})\left[
1-\tfrac{\tau-1}{\tau+1}(1-\alpha)\gamma_{\vc q} \right]+\epsilon \;.
\end{align}
Then the energy and intensity of the $T$ mode may be expressed as
\begin{equation}
\omega_{T\vc q} = \sqrt{a_{\vc q} b_{\vc q}} \;,\quad
I_{T\vc q} \propto \frac{g_{ab}^2}{\tau} \frac{\tau+1}2 
\sqrt{\frac{b_{\vc q}}{a_{\vc q}}}
\end{equation}
and for the $T'$ mode we have
\begin{equation}
\omega_{T'\vc q} = \omega_{T\vc{\tilde q}} \;,\quad
I_{T'\vc q} \propto \frac{g_c^2}{\tau} \frac{\tau-1}2 
\sqrt{\frac{a_{\vc{\tilde q}}}{b_{\vc{\tilde q}}}} \;,
\end{equation}
where $\vc{\tilde q} = \vc q + (\pi,\pi)$. 
As mentioned in the main text, the intensity contrast between the 
$T$ and $T'$ modes is most pronounced in the soft-spin situation
where $\tau$\,$\sim$\,1. At the crossing point of their dispersions, 
$\vc q$\,=\,$(\frac\pi2,\frac\pi2)$, $\gamma_{\vc q}$ is zero and $I_{T'}/I_T$
becomes $(g_c/g_{ab})^2 (\tau-1)/(\tau+1)$. For the relative intensity 
of the $L$ and $T$ mode at $\vc q$\,=\,$(0,0)$, used to quantify the proximity 
to the QCP, we get $I_L/I_T=2/\sqrt{(\tau+1)(\tau^2+1)}$ ($\alpha$\,=\,$0$, 
$\epsilon$\,=\,0) corrected by a multiplicative factor 
$1-\frac{\tau^2}{2(\tau+1)}\frac{\epsilon}{W}$ for small nonzero $\epsilon$.

The two-dimensional situation requires us to go beyond the harmonic
approximation for the amplitude mode. Its coupling to the two-magnon continuum
modifies the bare susceptibility
\begin{equation}
\chi_0(\vc q,\omega)=\frac{W}{2(\omega_{L\vc q}^2-\omega^2)}
\end{equation}
associated with the amplitude mode as $\chi^{-1}$\,=\,$\chi_0^{-1}-\Pi$. 
Collecting the leading terms, the self-energy $\Pi$ is obtained as
\begin{equation}
\Pi(\vc q,\omega) = \sum_{\vc k}
\frac{ M^2_{\vc k\vc k'} b_{\vc k}b_{\vc k'} 
(\omega^{-1}_{T\vc k}+\omega^{-1}_{T\vc k'}) }
{(\omega_{T\vc k}+\omega_{T\vc k'})^2-(\omega+i\Gamma)^2} \;.
\end{equation}
Here $\vc k'=-\vc k+\vc q+(\pi,\pi)$ and the matrix element
\begin{equation}
M^2_{\vc k\vc k'} = \frac{W^2}4 \left(1-\frac1{\tau^2}\right) 
\left( \frac{\gamma_{\vc q}}{\tau} 
      +\frac{\gamma_{\vc k}+\gamma_{\vc k'}}2 \right)^2 \;.
\end{equation}
In the calculations, we have used the broadening parameter $\Gamma$\,=\,$0.15W$.
The self-energy is largest for $\vc q$\,$\approx$\,$(\pi,\pi)$, where the dominant
contribution comes from $\vc k\approx -\vc k' \approx (\pi,\pi)$ (supported
by both small $\omega_T$ and large $M^2_{\vc k\vc k'}$), and turns 
the amplitude mode into a broad feature. 
A sizable gap $\omega_{T(\pi,\pi)}$ of the magnon dispersion prevents the infrared
singularity of $\Pi$, whose imaginary part would diverge like $1/\omega$ in
the gapless case. In our case it is zero below the cutoff energy 
$2\omega_{T(\pi,\pi)}$ comparable to $W$ making the above perturbative
approach well controlled.

\vspace{10 pt}
\noindent
{\bf C. Polarization analysis}

\begin{figure}
\renewcommand{\thefigure}{S\arabic{figure}}
  \centering
  \includegraphics[width=0.8\columnwidth]{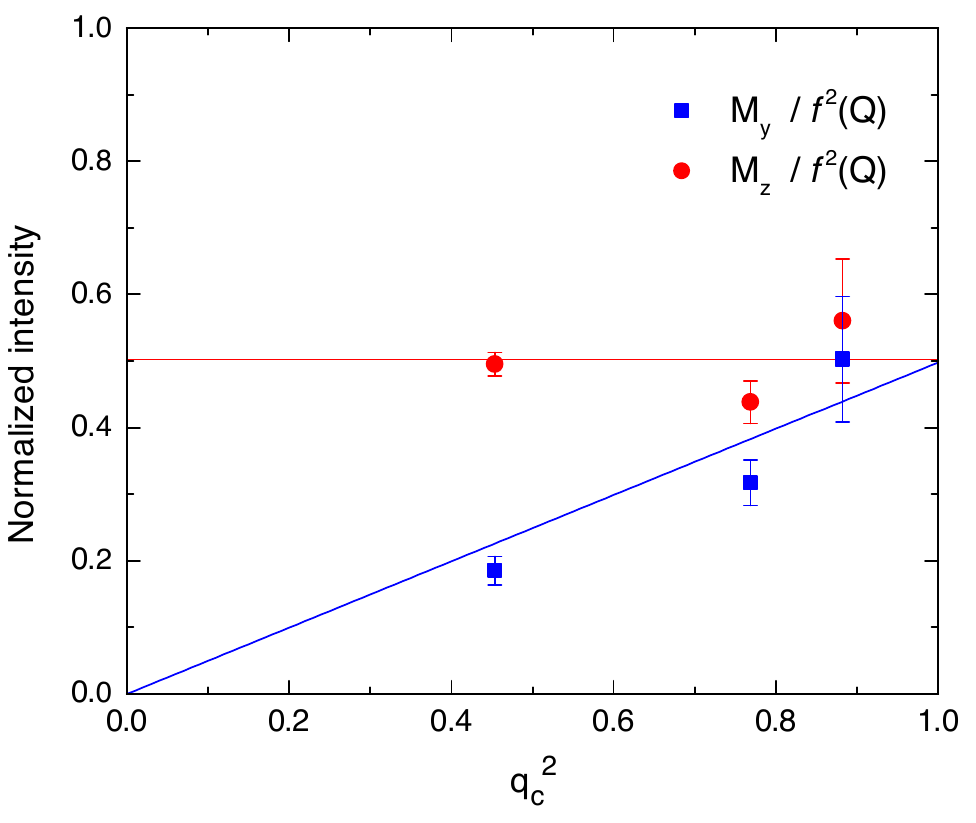}

  \caption{{\bf Determination of the twinning ratio.} Magnetic intensities $M_y$(blue squares) and $M_z$(red circles) normalized by the squared magnetic form factor as a function of $q_c^2$ at $\mathbf{Q}$\,=\,(1,0,$L$) with $L$\,=\,2, 4, and 6 at energy transfer 15\,meV. For one type of domain the intensity is constant, and for the other type of domain the intensity increases linearly with $q_c^2$. A one-parameter fit (red and blue solid lines) to the data points determines the twinning ratio $p$.}
\label{Fig:twin_ratio}
\end{figure}

\vspace{10 pt}
\noindent
In the standard reference frame for the neutron polarization with $\hat{x}\parallel\mathbf{Q}$, $\hat{y}\perp\mathbf{Q}$ in the scattering plane of the spectrometer and $\hat{z}=\hat{x}\times\hat{y}$, the magnetic intensity in the spin flip channels is extracted from the differences:
%
%
\begin{align}
  M_{y} &= I_x - I_y, \\ \nonumber
  M_{z} &= I_x - I_z,
\end{align}
where $I_x$, $I_y$, $I_z$ are the raw intensities of the respective polarizations. Note that any contribution from the background is suppressed in the difference. For conversion from INS intensity to dynamic spin susceptibility, we used the isotropic form factor for Ru$^+$, which gave a good description of the data at 15 meV (Fig.~\ref{Fig:twin_ratio}). 

\vspace{10 pt}
\noindent
{\bf C.1 Twinning ratio}

\vspace{10 pt}
\noindent
In this study the $a$ and $b$ orientation of the crystals in the array are not distinguished. In other words, for the volume fraction $p$ of the sample the scattering plane is ($H$,0,$L$), and for the fraction $(1-p)$ the scattering plane is (0,$H$,$L$). Taking into account the polarization factor, the intensities in each channel are related to excitations $M_{a}$, $M_{b}$ and $M_{c}$ along the crystallographic directions by:
\begin{align}
 M_{y} &= q_c^2 \left[ p~M_{a} + \left(1-p \right) M_{b} \right] + \left(1-q_c^2\right) M_{c}\\ \nonumber
 M_{z} &= \left(1-p \right) M_{a} + p~M_{b}
\end{align}
where $q_c^2$\,=\,$\left(Q_c/\lvert\mathbf{Q}\rvert\right)^2$.

The twinning ratio $p$ can be estimated from rocking scans through the Bragg reflections (4,0,0) and (0,4,0) where the separation in the scattering angle is large enough to distinguish the two peaks (not shown). Alternatively, $p$ can be estimated from the inelastic measurements by considering the $L$-dependence of the 15\,meV feature at ($\pi$,$\pi$) as shown in Fig.~\ref{Fig:twin_ratio}. Since this is an in-plane transverse mode, $M_b$ and $M_c$ vanish and eq.~(14) greatly simplifies. From the one-parameter fit to the data, a twinning ratio $p$\,=\,0.498$\,\pm\,0.014$ is determined, consistent with the first method. For the analysis we used $p$\,=\,0.5.  However, the conclusions drawn in the main text are unaffected by the exact value of $p$.

\vspace{10 pt}
\noindent
{\bf C.2 Energy scans at $\mathbf{q}$\,=\,(0,0)}
\vspace{10 pt}

\noindent
For the energy scans at $\mathbf{q}$\,=\,(0,0), it is useful to use two different Brillouin zones. The measurements at $\mathbf{Q}$\,=\,(2,0,0.4) at 15\,meV give conclusive evidence for the folded mode (Fig.~S3a), as the out-of-plane polarization gives rise to a signal in $M_y$ but not in $M_z$. We confirmed that the signal is peaked at (2,0,0.4) by scanning along the $H$ direction (not shown). To avoid a sharp spurion a small $L$ component was used. For the energy scan at $\mathbf{Q}$\,=\,(0,0,$L$), shown in Fig.~S3b, the signal exclusively originates from in-plane polarized modes. 
\begin{figure}
  \centering
  \includegraphics[width=0.95\columnwidth]{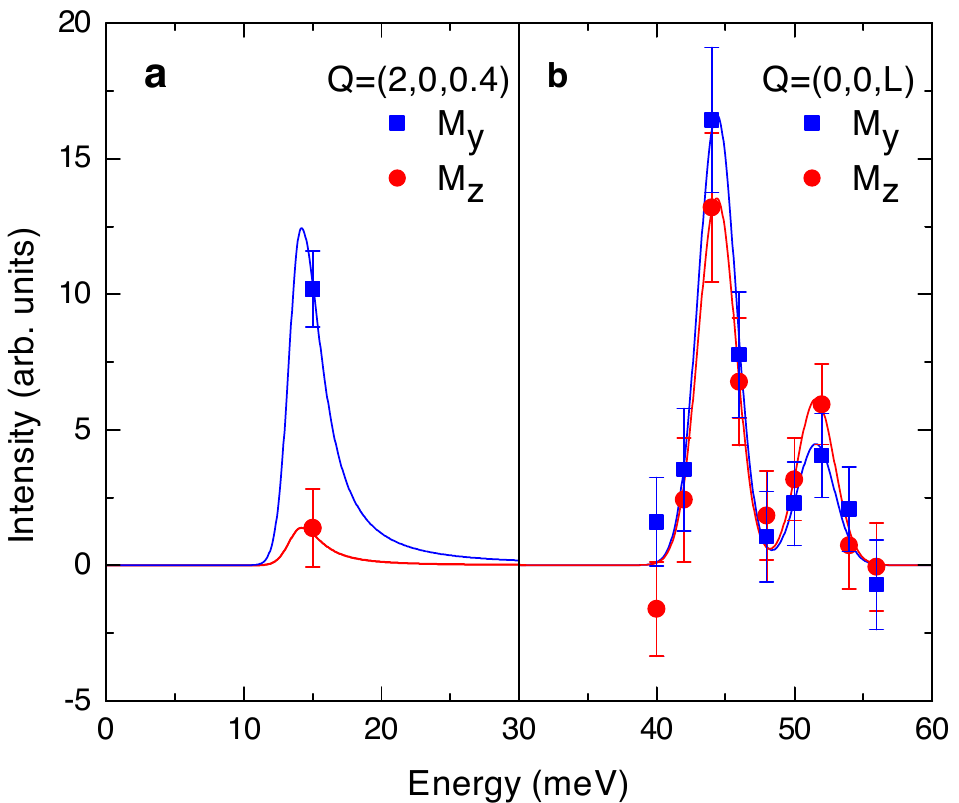}
\renewcommand{\thefigure}{S\arabic{figure}}
  \caption{{\bf Energy scans at $\mathbf{q}$\,=\,(0,0).} Magnetic intensities (a) at $\mathbf{Q}=(2,0,0.4)$, and (b) at $\mathbf{Q}=(0,0,L)$. The value of $L$ was varied along the scan to minimize the magnitude of $\mathbf{Q}$. Blue squares denote $M_y$, red circles $M_z$, and the lines are guides to the eye.}

\end{figure}

\begin{figure}
  \centering
  \includegraphics[width=0.95\columnwidth]{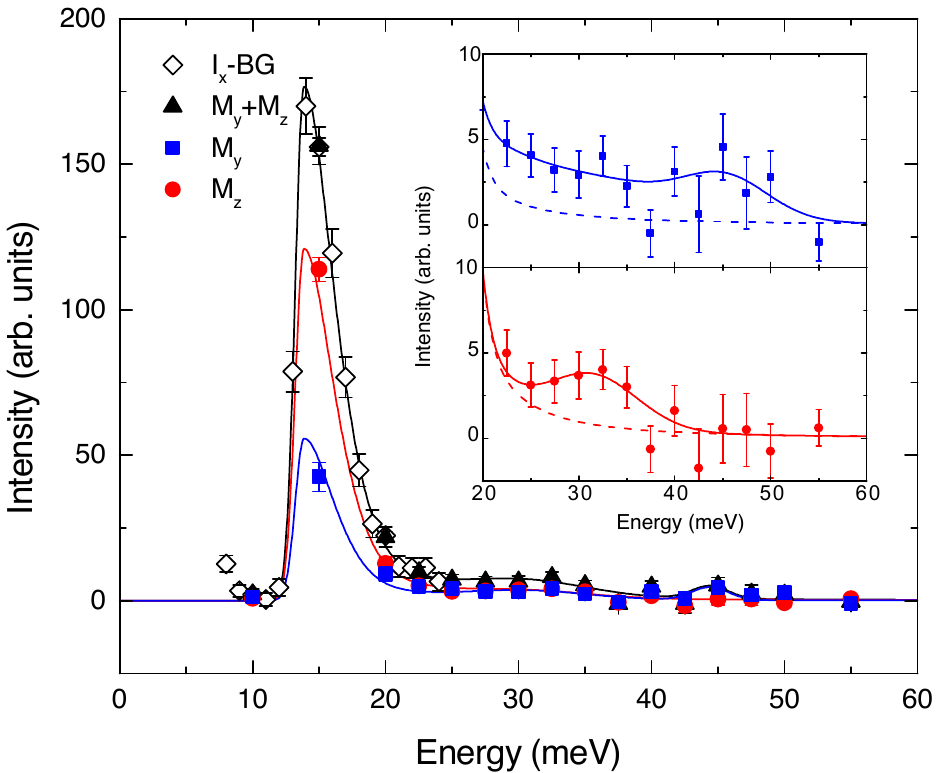}
\renewcommand{\thefigure}{S\arabic{figure}}
  \caption{{\bf Energy scan at $\mathbf{Q}$\,=\,(1,0,$L$).} The value of $L$ was varied along the scan to minimize the magnitude of $\mathbf{Q}$. The data denoted $I_x$-BG (empty black diamonds) is obtained from the raw data in the $M_x$ channel after subtraction of a small background; this method is only reliable when the signal is much larger than the background. The intensities $M_y$+$M_z$(filled black triangles), $M_y$ (blue squares), and $M_z$(red circles) are obtained using eq.~(13) and the lines are guides to the eye. The inset shows in detail the region above 20\,meV for $M_y$ (top) and $M_z$(bottom). Dashed lines represent the tail of the main transverse mode.}
\label{Fig:pipi}
\end{figure}

\vspace{10 pt}
\noindent
{\bf C.3 Energy scans at $\mathbf{q}$\,=\,$(\pi,\pi)$}
\vspace{10 pt}

\noindent
Energy scans at $\mathbf{Q}$\,=\,($1$,0,$L$) shown in Fig.~\ref{Fig:pipi}, corresponding to  $\mathbf{q}$\,=\,$(\pi,\pi)$ of the tetragonal unit cell, reveal three magnetic excitations above a gap of 14\,meV. For the two features lowest in energy we observe a signal in both $M_y$ and $M_z$ 
channels, characterizing them as in-plane polarized magnetic excitations. 
The third mode is unambiguously
identified as the folded mode as the polarization factor suppresses the intensity in the $M_{z}$ channel completely for out-of-plane excitations. The $M_y$ and $M_z$ signals allow separation of the in-plane and out-of-plane responses, because the $M_y$ signal exclusively originates from in-plane polarized modes, whereas the $M_z$ signal has contributions from both in-plane and out-of-plane polarized modes.

\null
\vfill

\vspace{20 pt}
\noindent
{\bf References}

\vspace{10 pt}
\noindent
$^{\mathrm{S1}}$Ulrich, C. \textit{et al.}, \textit{Phys. Rev. Lett.} \textbf{91}, 257202 (2003).

\vspace{4 pt}
\noindent
$^{\mathrm{S2}}$Khaliullin, G. \& Kilian, R. \textit{Phys. Rev. B} {\bf 61}, 3494 (2000).

\vspace{4 pt}
\noindent
$^{\mathrm{S3}}$Abragam, A \& Bleaney, B. {\it Electron Paramagnetic Resonance of Transition Ions} (Clarendon, Oxford, 1970).

\null
\vfill
\null
\vfill
\null
\vfill
\null
\vfill
\null
\vfill
\null
\vfill
\null
\vfill
\null
\vfill
\null
\vfill
\null
\vfill
\null
\vfill
\null
\vfill
\null
\vfill
\null
\vfill
\null
\vfill

\end{document}